\documentclass[%
print,
superscriptaddress,  
twocolumn,showpacs,amsmath,amssymb,9pt,
]{revtex4}

\usepackage{amsthm}
\usepackage{charter}
\usepackage{graphicx}
\usepackage{dcolumn}
\usepackage{bm}
\usepackage{dsfont}
\usepackage[landscape,papersize={297.1mm,210mm},left=1.6cm,right=1.46cm,top=2.3cm,bottom=2.56cm]{geometry}
\usepackage[                   
            pdfstartview=FitH,
            colorlinks, 
            pdfborder=001,   
            linkcolor=blue,
            anchorcolor=blue,
            citecolor=blue,
            urlcolor=blue
            ]{hyperref}

\begin{document}
\title{Monogamy of Logarithmic Negativity and Logarithmic Convex-Roof Extended Negativity}

\author{Li-Min Gao}
\affiliation {College of Physics, Hebei
Normal University, Shijiazhuang 050024, China}

\author{Feng-Li Yan}
\email{flyan@hebtu.edu.cn}
\affiliation {College of Physics, Hebei
Normal University, Shijiazhuang 050024, China}

\author{Ting Gao}
\email{gaoting@hebtu.edu.cn}
\affiliation {School of Mathematical Sciences, Hebei
Normal University, Shijiazhuang 050024, China}

\begin{abstract}
One of the fundamental traits of quantum entanglement is the restricted shareability among multipartite quantum systems, namely monogamy of entanglement, while it is well known that monogamy inequalities are always satisfied by entanglement measures with convexity. Here we present a measure of entanglement, logarithmic convex-roof extended negativity (LCREN) satisfying
important characteristics of an entanglement measure, and
investigate the monogamy relation for  logarithmic negativity and LCREN both without convexity. We show exactly that the $\alpha$th power of logarithmic negativity, and a newly defined good measure of entanglement, LCREN, obey a class of general monogamy inequalities in multiqubit systems, $2\otimes2\otimes3$ systems and $2\otimes2\otimes2^{n}$ systems for $\alpha\geq4\ln2$. We  provide a class of general polygamy inequalities of multiqubit systems in terms of logarithmic convex-roof extended negativity of assistance (LCRENoA) for $0\leq\beta\leq2$. Given that the logarithmic negativity and LCREN are not convex these results are surprising. Using the power of the logarithmic negativity and LCREN, we further establish a class of tight monogamy inequalities of multiqubit systems, $2\otimes2\otimes3$ systems and $2\otimes2\otimes2^{n}$ systems in terms of the $\alpha$th power of logarithmic negativity and LCREN for $\alpha\geq4\ln2$. We also show that the $\beta$th power of LCRENoA obeys a class of tight polygamy inequalities of multiqubit systems for $0\leq\beta\leq2$. \\

\end{abstract}


\maketitle

\maketitle

\emph{Introduction}.---A subsystem in a multipartite quantum system entangled with another subsystem limits its entanglement with the remaining ones. This behavior of entanglement is known as the monogamy of entanglement (MOE) [1, 2],  which means that entanglement cannot be freely shared unconditionally among the multipartite quantum systems, unlike classical correlations. For a three partite quantum system \emph{A}, \emph{B} and \emph{C}, in the clearest manifestation of MOE, if two systems \emph{A} and \emph{B} are maximally entangled, then neither of them can share any correlation --- let alone entanglement --- with a third party \emph{C}. This indicates that it should obey some trade-off on the amount of entanglement between the pairs \emph{AB} and \emph{AC}.

Since the monogamy of entanglement restricts on the amount of information that an eavesdropper could potentially obtain about the secret key extraction, it plays a crucial role in the context of quantum cryptography. Many information-theoretic protocols can be guaranteed secure by the MOE, such as quantum key distribution protocols [3-5]. MOE also has been used in many different fields of physics, such as quantum information theory [6], condensed-matter physics [7] and even black-hole physics [8].

MOE is one of the fundamental traits of entanglement and of quantum mechanics itself. The first quantitative characterization of the MOE was given by Coffman, Kundu and Wootters (CKW) for three-qubit state $\rho_{ABC}$ [1],
\begin{equation}\label{1}
E(\rho_{A|BC})\geq E(\rho_{AB})+ E(\rho_{AC}),
\end{equation}
where $E(\rho_{A|BC})=C^2(\rho_{A|BC})$ denotes the squared concurrence for quantifying bipartite entanglement between systems \emph{A} and \emph{BC} [9], $\rho_{AB}$ and $\rho_{AC}$ are reduced density matrices from a three-qubit state $\rho_{ABC}$. Since then, monogamy relation has been explored extensively [10-41]. Osborne and Verstraete presented a generalization of this three-qubit CKW monogamy inequality (1) to arbitrary multiqubit systems [10]. Later, the same monogamy inequality was generalized  in terms of various bipartite entanglement measures, such as the entanglement negativity and convex roof extended negativity \cite{14,15,16,17,18,19}, concurrence and entanglement of formation \cite{20, 21, 11}, Tsallis $q$-entropy and R\'{e}nyi-$\alpha$ entanglement \cite{22,23,24,25,26}, the one-way distillable entanglement and squashed entanglement \cite{27, 28}. Regula \textit{et al.} proposed a novel class of monogamy inequalities \cite{13}, which extended and sharpened the CKW inequality. Recently, some different kinds of monogamy relations for entanglement  were introduced \cite{34,35,36,38,39,40,41}. The monogamy relations involving other measures of quantum correlations were also given \cite{12,29,30,31,32,33}.

More recently, we find that measures of entanglement $E$ with monogamy property \cite{1, 10,11,12,13,14,15,16,17,18,19,20,21,22,23,24,25,26,27,28,35,36,41} are always convex. That is, whether a measure of entanglement without convexity obeys the monogamy inequality still remain unknown so far. As a well known measure of entanglement, the logarithmic negativity \cite{42, 43} is not convex. In addition, the logarithmic negativity, which possesses an operational interpretation \cite{44}, is an entanglement monotone both under general local operations and classical communication (LOCC) and positive partial transpose (PPT) preserving operations \cite{43}. The measure is the upper bound for distillable entanglement  \cite{42}, and is related to the entanglement cost under PPT preserving operations \cite{44}. Therefore the monogamy of logarithmic negativity is an important open question that needs to be settled.

In this Letter, we present a measure of entanglement, logarithmic convex-roof extended negativity (LCREN) satisfying
important characteristics of an entanglement measure, and provide a characterization of multipartite entanglement constraints in terms of logarithmic negativity, LCREN, and logarithmic convex-roof extended negativity of assistance (LCRENoA). By using the power of logarithmic negativity and LCREN, we establish a class of monogamy inequalities of multiqubit systems, $2\otimes2\otimes3$ systems and $2\otimes2\otimes2^{n}$ systems for $\alpha\geq4\ln2$. For $0\leq \beta \leq 2$, we establish a class of polygamy inequalities of multiqubit entanglement in terms of the $\beta$th power of LCRENoA. Given that the logarithmic negativity and LCREN are not convex these results are surprising, as it is generally considered that monogamy relation is satisfied by describing the local physical process of losing information. It is well known that tightening the monogamy and polygamy inequalities can provide a precise characterization of the entanglement sharing and distribution in multipartite systems. We further present a class of tight monogamy inequalities in terms of logarithmic negativity and LCREN for multiqubit systems, $2\otimes2\otimes3$ systems and $2\otimes2\otimes2^{n}$ systems, and a class of tight polygamy inequalities in terms of LCRENoA for multiqubit systems.

\emph{The general monogamy inequalities for logarithmic negativity and LCREN, and general polygamy inequalities for LCRENoA}.---Next, we provide a class of monogamy inequalities
of multipartite entanglement using the power of logarithmic negativity  and LCREN,  and polygamy inequalities for LCRENoA. Before we present our main results, we first provide some notations, definitions and lemmas, which are useful throughout this paper.

For a quantum state $\rho_{AB}$ on Hilbert space $\mathcal{H}_{A}\otimes \mathcal{H}_{B}$, its negativity, $\mathcal{N}(\rho_{AB})$ is defined as \cite{42, 43, 45}
\begin{equation}\label{2}
\mathcal{N}(\rho_{AB})=\|\rho_{AB}^{T_A}\|_{1}-1,
\end{equation}
where $\rho_{AB}^{T_A}$ denotes the partial transpose of $\rho_{AB}$ with respect to the subsystem $A$, and the trace norm $\|X\|_{1}=\text{tr}\sqrt{X X^\dag}$.

For an arbitrary $N$-qubit pure quantum state $|\psi\rangle_{AB_{1}\cdots B_{N-1}}$, $|\psi\rangle_{A|B_{1}\cdots B_{N-1}}$ denotes the state $|\psi\rangle_{AB_{1}\cdots B_{N-1}}$ viewed as a bipartite state under the partition $A$ and $B_{1}B_{2}\cdots B_{N-1}$. Then, the negativity $\mathcal{N}(|\psi\rangle_{A|B_{1}\cdots B_{N-1}})$ satisfies \cite{14}
\begin{equation}\label{3}
 \begin{aligned}
\mathcal{N}^{2}(|\psi\rangle_{A|B_{1}\cdots B_{N-1}})\geq \sum\limits_{i=1}^{N-1}\mathcal{N}^{2}(\rho_{AB_{i}}),
\end{aligned}
\end{equation}
where the reduced density matrices $\rho_{AB_{i}}$ are obtained by tracing over the subsystems $B_1$, ..., $B_{i-1}, B_{i+1}$, ..., $B_{N-1}$.

A more easily interpreted and computable measure of entanglement is the logarithmic negativity, which is defined as \cite{42, 43}
\begin{equation}\label{4}
E_{\mathcal{N}}(\rho_{AB})=\log_{2}\|\rho_{AB}^{T_A}\|_{1}=\log_{2}[\mathcal{N}(\rho_{AB})+1].
\end{equation}
This quantity is an entanglement monotone both under general LOCC and PPT preserving operations but not convex \cite{43}. It is, moreover, additive.

By construction, the negativity fails to recognize entanglement in PPT states. In order to overcome its lack of separability criterion, one modification of negativity  is convex-roof extended negativity (CREN), which gives a perfect
discrimination of PPT bound entangled states and separable states in any bipartite quantum system.

For a bipartite  state $\rho_{AB}$, its CREN, $\mathcal{\widetilde{N}}(\rho_{AB})$, is defined by  \cite{46}
\begin{equation}\label{5}
\mathcal{\widetilde{N}}(\rho_{AB})=\min_{\{p_{k}, |\varphi_{k}\rangle_{AB}\}}\sum_{k}{p_{k}}{\mathcal{N}}(|\varphi_{k}\rangle_{AB}),
\end{equation}
while the CREN of assistance (CRENoA), which can be considered
to be dual to CREN,  is defined  as \cite{18}
\begin{equation}\label{6}
\mathcal{\widetilde{N}}_a(\rho_{AB})=\max_{\{p_{k}, |\varphi_{k}\rangle_{AB}\}}\sum_{k}{p_{k}}{\mathcal{N}}(|\varphi_{k}\rangle_{AB}),
\end{equation}
where the minimum and maximum are taken over all possible pure-state decompositions of $\rho_{AB}=\sum_{k}p_{k}|\varphi_{k}\rangle_{AB}\langle\varphi_{k}|$.
By definition, both the CREN and CRENoA of a pure state are equal to its negativity.

The convexity of negativity, and Eqs. (\ref{5})-(\ref{6}) result in
\begin{equation}\label{N-CREN-CRENoA}
 {\mathcal{N}}(\rho_{AB})\leq {\mathcal{\widetilde{N}}}(\rho_{AB})\leq {\mathcal{\widetilde{N}}_a}(\rho_{AB}).
\end{equation}

For an arbitrary $N$-qubit state $\rho_{AB_{1}\cdots B_{N-1}}$ and its reduced density matrices $\rho_{AB_i}$,  the square of CREN  satisfies the following monogamy inequality \cite{18,19}
\begin{equation}\label{7}
 \begin{aligned}
 \mathcal{\widetilde{N}}^{2}(\rho_{A|B_{1}\cdots B_{N-1}})\geq \sum\limits_{i=1}^{N-1} \mathcal{\widetilde{N}}^{2}(\rho_{AB_{i}}).
\end{aligned}
\end{equation}
For any pure state $|\psi\rangle_{AB_{1}\cdots B_{N-1}}$ of $N$-qubit systems, the following polygamy inequality  \cite{18} holds
\begin{equation}\label{8}
 \begin{aligned}
 \mathcal{\widetilde{N}}^{2}_a(|\psi\rangle_{A|B_{1}\cdots B_{N-1}}) \leq  \sum\limits_{i=1}^{N-1} \mathcal{\widetilde{N}}^{2}_a(\rho_{AB_{i}}).
\end{aligned}
\end{equation}

For any bipartite state $\rho_{AB}$, we define LCREN as
\begin{equation}\label{9}
E_{\mathcal{\widetilde{N}}}(\rho_{AB})=\log_{2}[\mathcal{\widetilde{N}}(\rho_{AB})+1].
\end{equation}
Clearly, LCREN is invariant under local unitary transformations. One important property is this: $E_{\mathcal{\widetilde{N}}}(\rho_{AB})$ is nonzero if
and only if $\rho_{AB}$ is entangled (and so it equals zero if
and only if $\rho_{AB}$ is separable). Besides, it is entanglement monotone under LOCC operations. LCREN is not only nonincreasing under LOCC, but also nonincreasing on average under
LOCC, which follow from the entanglement monotonicity of CREN under LOCC, the monotonicity logarithm, and concavity of logarithm.

However, just as logarithmic negativity, LCREN is also not convex. Suppose that  $\rho_{AB}=\sum_{k}p_{k}\rho_k$ with $\rho_k=|\varphi_k\rangle_{AB}\langle\varphi_k|$ is the optimal decomposition for $\rho_{AB}$
achieving the minimum of (\ref{5}). Then  $\mathcal{\widetilde{N}}(\rho_{AB})= \sum_{k}{p_{k}}{\mathcal{N}}(|\varphi_{k}\rangle_{AB})$ by definition.  The concavity of logarithm ensures
\begin{equation}\label{LCREN not convex}
\begin{array}{cl}
  E_{\mathcal{\widetilde{N}}}(\sum_{k}p_{k}\rho_k) & =\log_2[\sum_{k}{p_{k}}{\mathcal{N}}(|\varphi_{k}\rangle_{AB})+1] \\
  & =\log_2[\sum_{k}{p_{k}}\|\rho_k^{T_A}\|_1] \\
   & \geq \sum_{k}{p_{k}} \log_2\|\rho_k^{T_A}\|_1\\
  & = \sum_{k}p_{k} E_{\mathcal{\widetilde{N}}}(\rho_k),
\end{array}
\end{equation}
which implies that LCREN is not convex.

Similar to the duality between CREN and CRENoA, we can also define a dual to LCREN, namely LCRENoA, by
\begin{equation}\label{10}
E_{\mathcal{\widetilde{N}}_a}(\rho_{AB})=\log_{2}[\mathcal{\widetilde{N}}_a(\rho_{AB})+1].
\end{equation}

By the monotonicity of  logarithm, and Eq.(\ref{N-CREN-CRENoA}), we arrive at
\begin{equation}\label{11}
E_{\mathcal{N}}(\rho_{AB})\leq E_{\mathcal{\widetilde{N}}}(\rho_{AB})\leq E_{\mathcal{\widetilde{N}}_a}(\rho_{AB}).
\end{equation}

In order to investigate the general monogamy inequality for logarithmic negativity and LCREN, and the polygamy for LCRENoA, we need the following lemmas, whose analytical proofs can be found in the Supplemental Material \cite{47}.

\emph{Lemma 1}.  For $\alpha\geq4\ln2$, and $0\leq\beta\leq2$, there are
\begin{equation}\label{12}
[\log_{2}(1+\sqrt{x^{2}+y^{2}})]^{\alpha}\geq [\log_{2}(1+x)]^{\alpha}+[\log_{2}(1+y)]^{\alpha},
\end{equation}
and
\begin{equation}\label{13}
[\log_{2}(1+\sqrt{x^{2}+y^{2}})]^{\beta}\leq [\log_{2}(1+x)]^{\beta}+[\log_{2}(1+y)]^{\beta}
\end{equation}
on the domain $D=\{(x,y)|0\leq x,y,x^{2}+y^{2}\leq1\}$.

\emph{Lemma 2}. For any pure quantum state $|\phi\rangle_{ABC}$ of $2\otimes2\otimes3$ systems, we have
\begin{equation}\label{14}
  \mathcal{N}^{2}(|\phi\rangle_{A|BC})  \geq  \mathcal{N}^{2}(\rho_{AB})+\mathcal{N}^{2}(\rho_{AC}),
\end{equation}
while for any pure quantum state $|\phi\rangle_{ABC}$ of $2\otimes2\otimes3$ systems, or $2\otimes2\otimes 2^n$ systems, we have
\begin{equation}\label{15}
 \mathcal{\widetilde{N}}^{2}(|\phi\rangle_{A|BC}) \geq  \mathcal{\widetilde{N}}^{2}(\rho_{AB})+\mathcal{\widetilde{N}}^{2}(\rho_{AC}).
\end{equation}

Inequalities (\ref{14})  also hold for any pure state of $2\otimes2\otimes2^n$ systems \cite{19}.

Now, we are ready to have the following theorems, which states that a class of monogamy and polygamy inequalities of multipartite entanglement can be established using the power logarithmic nengativity, LCREN, and LCRENoA.

\emph{Theorem 1}.  For any $N$-qubit pure state $|\psi\rangle_{A B_{1}\cdots B_{N-1}}$, we have monogamy inequality
\begin{equation}\label{16}
 \begin{aligned}
E_{\mathcal{N}}^{\alpha}(|\psi\rangle_{A|B_{1}\cdots B_{N-1}})\geq \sum\limits_{i=1}^{N-1}  E_{\mathcal{N}}^{\alpha}(\rho_{AB_{i}}),
\end{aligned}
\end{equation}
for $\alpha\geq4\ln2$, and polygamy inequality
\begin{equation}\label{22}
 \begin{aligned}
E_{\mathcal{\widetilde{N}}_a}^{\beta}(|\psi\rangle_{A|B_{1}\cdots B_{N-1}})
 \leq \sum\limits_{i=1}^{N-1} E_{\mathcal{\widetilde{N}}_a}^{\beta}(\rho_{AB_{i}}),
\end{aligned}
\end{equation}
for $0\leq\beta\leq2$, while  for an arbitrary $N$-qubit state $\rho_{AB_{1}\cdots B_{N-1}}$ and $\alpha\geq4\ln2$, we have monogamy inequality
\begin{equation}\label{20}
 \begin{aligned}
E_{\mathcal{\widetilde{N}}}^{\alpha}(\rho_{A|B_{1}\cdots B_{N-1}})
 \geq \sum\limits_{i=1}^{N-1}  E_{\mathcal{\widetilde{N}}}^{\alpha}(\rho_{AB_{i}}).
\end{aligned}
\end{equation}

\emph{Proof}. Employing inequality (\ref{3}), the monotonicity of logarithm,  and the inequality (\ref{12}), we find
\begin{equation}\label{17}
 \begin{aligned}
 & E_{\mathcal{N}}^{\alpha}(|\psi\rangle_{A|B_{1}\cdots B_{N-1}})\\
  = & \left(\log_{2}\left[\mathcal{N}(|\psi\rangle_{A|B_{1}\cdots B_{N-1}})+1\right]\right)^{\alpha}\\
  \geq & \left(\log_{2}\left[\sqrt{\sum\limits_{i=1}^{N-1}\mathcal{N}^{2}(\rho_{AB_{i}})}+1\right]\right)^{\alpha}\\
  \geq & \left(\log_{2}[\mathcal{N}(\rho_{AB_{1}})+1]\right)^{\alpha}+\left(\log_{2}\left[\sqrt{\sum\limits_{i=2}^{N-1}\mathcal{N}^{2}(\rho_{AB_{i}})}+1\right]\right)^{\alpha}\\
   \geq &\left(\log_{2}[\mathcal{N}(\rho_{A|B_{1}})+1]\right)^{\alpha}+\left(\log_{2}[\mathcal{N}(\rho_{A|B_{2}})
  +1]\right)^{\alpha}\\ \quad & +\cdots+\left(\log_{2}[\mathcal{N}(\rho_{A|B_{N-1}})+1]\right)^{\alpha}\\
  = & \sum\limits_{i=1}^{N-1}  E_{\mathcal{N}}^{\alpha}(\rho_{AB_{i}}),
\end{aligned}
\end{equation}
where we have utilized the monogamy inequality (\ref{3}) and the monotonically increasing property of logarithm in the first inequality, while the second inequality is due to inequality (\ref{12}) by letting $x=\mathcal{N}(\rho_{AB_{1}})$ and $y=\sqrt{\mathcal{N}^{2}(\rho_{AB_{2}})+\cdots+\mathcal{N}^{2}(\rho_{AB_{N-1}})}$. The third inequality is obtained from the iterative use of inequality (\ref{12}). This completes the proof of (\ref{16}).

Similarly, inequality (\ref{22}) follows from inequality (\ref{8}), the monotonicity of logarithm, and the iterative use of inequality (\ref{13}), while (\ref{20}) follows from Eq.(\ref{7}), the monotonicity of logarithm, and Eq.(\ref{12}).

\emph{Theorem 2}. For any pure state $|\phi\rangle_{ABC}$ of $2\otimes2\otimes3$ or  $2\otimes2\otimes2^n$ systems,  we have
\begin{eqnarray}
 E_{\mathcal{N}}^{\alpha}(|\phi\rangle_{A|BC}) & \geq & E_{\mathcal{N}}^{\alpha}(\rho_{AB})+ E_{\mathcal{N}}^{\alpha}(\rho_{AC}),\label{18}\\
 E_{\mathcal{\widetilde{N}}}^{\alpha}(|\phi\rangle_{A|BC}) & \geq & E_{\mathcal{\widetilde{N}}}^{\alpha}(\rho_{AB})+ E_{\mathcal{\widetilde{N}}}^{\alpha}(\rho_{AC}),\label{19}
\end{eqnarray}
while for any mixed tripartite state  $\rho_{ABC}$  of $2\otimes2\otimes2^n$ systems, we have
 \begin{equation}\label{21}
 \begin{aligned}
 E_{\mathcal{\widetilde{N}}}^{\alpha}(\rho_{A|BC})\geq E_{\mathcal{\widetilde{N}}}^{\alpha}(\rho_{AB})+ E_{\mathcal{\widetilde{N}}}^{\alpha}(\rho_{AC}).
\end{aligned}
\end{equation}
Here  $\alpha\geq4\ln2$.

Inequalities (\ref{18}) and (\ref{19}) follow immediately from Lemma 2, the monotonicity of logarithm, and the inequality (\ref{12}).  Note that
\begin{eqnarray} \label{2-2-2n}
  \mathcal{\widetilde{N}}^{2}(\rho_{A|BC}) & \geq & \mathcal{\widetilde{N}}^{2}(\rho_{AB})+\mathcal{\widetilde{N}}^{2}(\rho_{AC})
\end{eqnarray}
hold for any tripartite state of $2\otimes2\otimes2^n$ systems \cite{19}. Then by the monotonicity of logarithm, and the inequality (\ref{12}), one has (\ref{21}).

Both logarithmic negativity and LCREN are good measures of entanglement in multipartite systems for MOE, while LCRENoA is a good measure of entanglement in multiqubit systems for polygamy of entanglement. These monogamy and polygamy inequalities above can be further refined and become tighter. In fact, Theorems 1 and 2 can even be improved to be tighter inequalities with some condition on the logarithmic negativity, LCREN, and LCRENoA. The results and the proofs are given in the Supplemental Material \cite{47}.

\emph{Conclusion}.---The logarithmic negativity is an entanglement monotone both under general LOCC and PPT preserving operations. Importantly, this measure possesses an operational interpretation and is easy to calculate. It is therefore an important task to investigate the monogamy of logarithmic negativity. The newly defined measure, LCREN, as well as logarithmic negativity is  an good entanglement measure without convexity. We provide a characterization of multipartite entanglement constraints in terms of logarithmic negativity, LCREN, and LCRENoA. We have established a class of monogamy inequalities in multiqubit systems, $2\otimes2\otimes3$ systems and $2\otimes2\otimes2^{n}$ systems based on the $\alpha$th power of logarithmic negativity and LCREN for $\alpha\geq4\ln2$. Given that the logarithmic negativity and LCREN are not convex these results are surprising, as it is generally considered that monogamy inequalities are always satisfied by entanglement measures with convexity. We also show that the $\alpha$th power of logarithmic negativity and LCREN provide a class of monogamy inequalities of multiqubit systems, $2\otimes2\otimes3$ systems and $2\otimes2\otimes2^{n}$ systems entanglement in a tight way for $\alpha\geq4\ln2$. We further provides the general and tight polygamy inequalities of multiqubit systems using the $\beta$th power of LCRENoA for $0\leq\beta\leq2$. These results provide finer characterizations of multipartite quantum entanglement sharing and distribution among the multipartite systems. Given the importance of the study on multipartite quantum entanglement, our results not only can provide a rich reference for future work on the study of multipartite quantum entanglement, but also may contribute to a fully understanding of the multipartite quantum entanglement.

\vspace{0.6cm}
\acknowledgments
This work was supported by the National Natural Science Foundation of China under Grant No: 11475054, the Hebei Natural Science Foundation of China under Grant Nos. A2020205014, A2018205125, and the Education Department of Hebei Province Natural Science Foundation under Grant No. ZD2020167.

\end{document}